\documentclass[prb,floatfix,twocolumn,showpacs,amsmath,amssymb]{revtex4}
 \usepackage{graphicx}
\usepackage{dcolumn}
\usepackage{bm}
\usepackage[usenames]{color}
\newcommand{\dagga}{{\phantom{\dagger}}}
\begin{document}

\title{Bose-glass, superfluid, and rung-Mott phases of hard-core bosons in 
disordered two-leg ladders}
\author{Juan Carrasquilla,$^{1,2}$ Federico Becca,$^{2,3}$ and Michele Fabrizio$^{2,3,4}$}
\affiliation{
$^{1}$ Department of Physics, Georgetown University, Washington, DC 200057, USA \\
$^{2}$ International School for Advanced Studies (SISSA), Via Beirut 2, I-34151, Trieste, Italy \\
$^{3}$ Democritos Simulation Center CNR-IOM Istituto Officina dei Materiali, Trieste, Italy \\
$^{4}$ International Centre for Theoretical Physics (ICTP), P.O. Box 586, I-34014 Trieste, Italy}
\date{\today}

\begin{abstract}
By means of Monte Carlo techniques, we study the role of disorder on a system 
of hard-core bosons in a two-leg ladder with both intra-chain ($t$) and 
inter-chain ($t^\prime$) hoppings. We find that the phase diagram as a function
of the boson density, disorder strength, and $t^\prime/t$ is far from being 
trivial. This contrasts the case of spin-less fermions where standard 
localization arguments apply and an Anderson-localized phase pervades the whole
phase diagram. A compressible Bose-glass phase always intrudes between the Mott 
insulator with zero (or one) bosons per site and the superfluid that is 
stabilized for weak disorder. At half filling, there is a direct transition 
between a (gapped) rung-Mott insulator and a Bose glass, which is driven by 
exponentially rare regions where disorder is suppressed. Finally, by doping the
rung-Mott insulator, a direct transition to the superfluid is possible only in 
the clean system, whereas the Mott phase is always surrounded by the a Bose 
glass when disorder is present. The phase diagram based on our numerical 
evidence is finally reported.
\end{abstract}

\pacs{05.30.Jp, 71.27.+a,71.30.+h}

\maketitle

\section{Introduction}

Interacting bosons in one-dimensional (1D) or quasi-1D lattices are of interest
in many physical contexts, ranging from Josephson-junction arrays~\cite{fazio} 
to more recent experiments on ultracold bosons loaded in 1D optical 
traps.~\cite{stoferle} Especially the latter ones offer the unique opportunity
to fine tune the experimental parameters and realize in laboratory a wide 
variety of bosonic models where kinetic energy, inter-particle interaction and
disorder can be varied at will. In particular, it is possible to tune 
interaction to such an extend that atoms essentially behave as hard-core bosons
confined along 1D tubes with~\cite{paredes} and without~\cite{kinoshita} 
a superimposed optical lattice. In this condition, the huge repulsion prevents
bosons from occupying the same position in space and induces a sort of Pauli 
exclusion principle. It is well known that hard-core bosons can be mapped onto
spin-less fermions by the so-called Jordan-Wigner transformation.~\cite{jordan}
However, since the Jordan-Wigner fermions are non local in terms of the 
original bosonic operators, the local bosonic Hamiltonian is transformed into 
a very complicated non-local interacting fermion model. This approach 
simplifies in 1D tight-binding models with only nearest-neighbor hopping: 
here, non-locality is absent and the bosonic model maps onto non-interacting 
fermions, easily solvable. Whenever the lattice is not rigorously 1D, i.e., 
when more chains are coupled together or longer-range hoppings are considered,
the corresponding fermionic problem contains complicated interaction terms 
that may become highly non-local when the number of chains increases.

The difference between hard-core bosons and spin-less fermions is even more 
pronounced in the presence of disorder. Indeed, non-interacting spin-less 
fermions in a quasi-1D system are always Anderson localized for any disorder 
strength. On the contrary, hard-core bosons can become superfluid in the 
presence of disorder as soon as they can exchange among each other, a property
that occurs already in the simplest case of a two-leg 
ladder.~\cite{orignac,orignac2} From this point of view, ladders of hard-core 
bosons represent an ideal case study to uncover the role of Bose statistics 
versus Fermi statistics in the presence of disorder. 

From the purely theoretical side, we think this is an interesting issue. 
Single-particle wave functions are always localized in a quasi-1D disordered 
lattice. It follows that any Slater determinant built with such wave functions
is localized too, so any many-body wave function for non-interacting fermions. 
On the contrary, hard-core bosons, which like spin-less fermions cannot occupy
the same site but whose wave function is symmetric under exchanging two 
particles, can cooperatively act and give rise to a delocalized superfluid 
phase.     

Also from the experimental side ladder systems are of interest, as they 
can be realized with optical lattices.~\cite{clark,clarkpra} In realistic 
experimental setups, a two-leg ladder can be realized through a double-well
potential along a direction (say, $y$) like in Ref.~\onlinecite{oberthaler}, 
and a potential creating a cigar geometry in the $x$-axis. Further 
superimposing a periodic potential along $x$, one could finally realize a 
two-leg Bose-Hubbard model with tunable hopping rates among and between the 
legs. Disorder can be introduced by superimposing a disordering lattice or 
introducing a speckle potential. We further mention that bosonic ladder systems
are realized also in magnetic materials.~\cite{tchernyshyov} For example, the 
disorder-free compound IPA-CuCl$_3$ has been found to be a prototypical 
$S=1/2$ antiferromagnetic spin ladder material, which can be thought as a 
system of interacting hard-core bosons.~\cite{masuda} Here disorder is 
introduced by means of random chemical substitution, i.e., 
IPA-Cu(Cl$_{1-x}$Br$_x$)$_3$. Neutron scattering experiments have shown 
convincing evidences of the spin-analogous of a Bose-glass.~\cite{hong}

Given its both theoretical and experimental interest, we decided to study 
the phase diagram of a simple model of hard-core bosons hopping on a two-leg 
ladder with bounded on-site disorder by means of Green's function Monte 
Carlo.~\cite{trivedi} In short, we find that the phase diagram as function of 
the density and the ratio between inter- and intra-chain hopping includes
three phases: a localized Bose-glass, a superfluid and, at half-filling, a 
so-called rung-Mott insulator that seems to be always surrounded by the glass.

The paper is organized as follows: in Sec.~\ref{sec:model}, we introduce the 
model and the numerical methods; in Sec.~\ref{sec:clean}, we briefly discuss 
the clean system; in Sec.~\ref{sec:disordered}, we present our results for 
the disordered model; in Sec.~\ref{sec:concl}, we finally draw our conclusions.
 
\section{The Model}\label{sec:model}

We shall consider a system of disordered hard-core bosons on a $L=2 \times L_x$ 
lattice. The Hamiltonian reads 
\begin{eqnarray}\label{hcbosons}
{\cal H} = &-& t \sum_{i, \eta=1,2 } 
\left( b^\dagger_{i,\eta} b^\dagga_{i+1,\eta} + h.c.\right) 
- t^\prime \sum_{i} \left( b^\dagger_{i,1} b^\dagga_{i,2} + h.c.\right) \nonumber \\
&+& \sum_{i,\eta} \epsilon_{i,\eta} n_{i,\eta},
\end{eqnarray}
where $b^\dagger_{i,\eta}$ ($b^\dagga_{i,\eta}$) creates (destroys) a boson 
at rung $i$ on the chain $\eta=1,2$. The matrix elements $t$ and $t^\prime$ 
are the hopping amplitudes along legs and rungs, respectively. The disordered 
potential couples to the density operator 
$n_{i,\eta}=b^\dagger_{i,\eta} b_{i,\eta}$ and it is described by random 
variables $\epsilon_{i,\eta}$ that are uniformly distributed in 
$[-\Delta,\Delta]$. Finally, the hard-core constraint of $n_{i,\eta} \le 1$ 
is implied.

We study the Hamiltonian Eq.~(\ref{hcbosons}) by Green's function Monte 
Carlo with a fixed number $M$ of bosons on $L$ sites,~\cite{calandra} $n=M/L$ 
being the average density. The Green's function Monte Carlo approach is based 
on a stochastic implementation of the power method technique that allows, 
in principle, to extract the actual ground state $|\Psi_{GS}\rangle$ of a 
given Hamiltonian ${\cal H}$, from any starting trial (e.g. variational) 
wave function $|\Psi_{V} \rangle$, provided that 
$\langle \Psi_{V} | \Psi_{GS}\rangle \ne 0 $. In order to improve the numerical 
efficiency, it is important to consider a good starting wave function, for 
which we use one- and two-body Jastrow factors applied to a state where all 
bosons are condensed at momentum $q=0$. The one-body Jastrow factor makes it
possible to vary the local density of the bosons. On the contrary, the 
two-body Jastrow is taken to be translationally invariant.~\cite{carrasqu}

\begin{figure}
\includegraphics[width=\columnwidth]{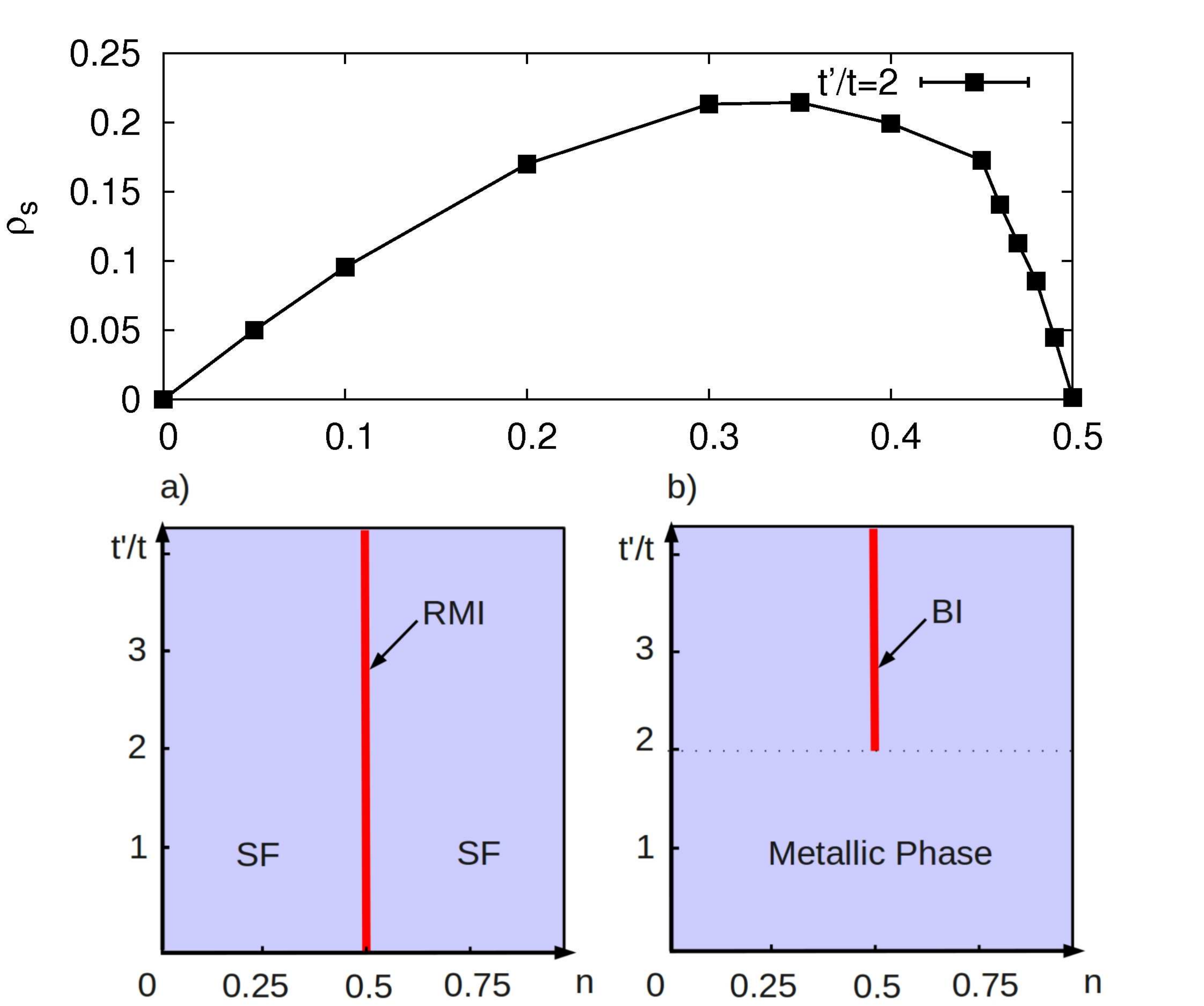}
\caption{\label{fig:stiffclean} 
(Color on-line) Upper panel: Superfluid stiffness as function of the density 
$n$ at fixed $t^\prime/t=2$ on a lattice with $L=2 \times 50$. Lower panel: 
clean phase diagram of hard-core bosons (a) and spin-less fermions (b) on the
two-leg ladder as function of $t^\prime/t$ and density $n$. The rung-Mott 
insulator is denoted by RMI, the superfluid by SF and the band insulator by 
BI.}
\end{figure}

\section{The clean system}\label{sec:clean}

Before considering the disordered case, it is useful to briefly discuss the 
clean system, where $\epsilon_{i,\eta}=0$ (see also Appendix). First, we 
consider the limit $t^\prime=0$, i.e., two uncoupled chains. In this situation,
the ground state is a superfluid with quasi-long-range order for any density 
$0<n<1$. At densities $n=0$ and $1$, there is a (trivial) ``frozen'' Mott 
insulator due to the infinite on-site repulsion, which completely suppresses 
charge fluctuations. Let us now analyze the the opposite limit 
$t^\prime/t \gg 1$. Exactly at half filling, i.e., $n=0.5$, there is one boson
per rung and the wave function can be approximately written as a independent 
product of single-particle rung states as
\begin{equation}\label{largetprime}
|\Psi_{GS}\rangle\simeq \prod_i^{L_x} 
(b^\dagger_{i,1} + b^\dagger_{i,2})|0\rangle.
\end{equation}
The system is in the so-called rung-Mott insulator with a unique ground state 
and a gap to all excitations.~\cite{fisherrung} At half filling, the transition
between the rung-Mott insulator and the superfluid takes place exactly at
$t^\prime=0$, since the inter-chain hopping represents a relevant perturbation
that immediately opens a gap in the excitation spectrum, see Appendix. 
This transition is of the Berezinskii-Kosterlitz-Thouless type, which makes 
it difficult to observe by numerical simulations on finite 
clusters.~\cite{nicolas} 

Now, if few bosons are added or removed to the insulating state, a superfluid 
is stabilized. Therefore, at any other filling $0<n<0.5$ and $0.5<n<1$ the 
system will be always superfluid, as can be seen in Fig.~\ref{fig:stiffclean} 
where the superfluid stiffness as function of the density of particles $n$ is 
shown for fixed $t^\prime/t=2$. The superfluid stiffness starts to grow at low
density, it reaches a maximum, and finally vanishes at $n=0.5$ at the 
rung-Mott insulator. 

We conclude this section by mentioning that, for spin-less fermions at half
filling, there is a transition from a metallic to a band insulator at 
$t^\prime/t=2$, where a gap opens up. For all the other densities $n \ne 0$ 
and $1$, the ground state is metallic. The clean phase diagrams for 
hard-core bosons and spin-less fermions on a two-leg ladder are sketched 
in Fig.~\ref{fig:stiffclean} for comparison.

\begin{figure}
\includegraphics[width=\columnwidth]{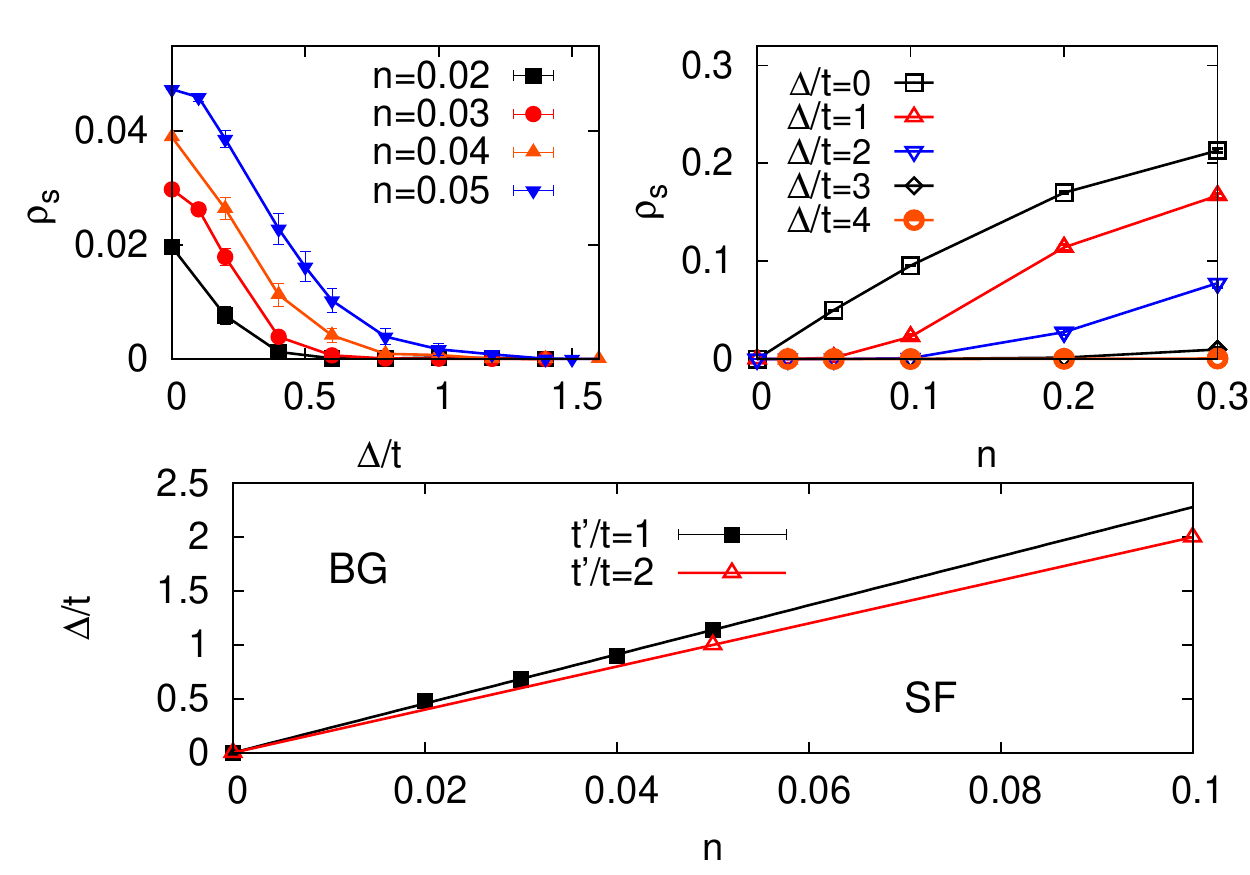}
\caption{\label{fig:hcb}
(Color on-line) Upper left panel: superfluid stiffness $\rho_s$ as a function 
of the disorder strength $\Delta/t$ for different $n$ and $t^\prime/t=1$. 
Upper right panel: superfluid stiffness $\rho_s$ as a function of the density 
for different $\Delta/t$ and fixed $t^\prime/t=2$. Lower panel: low-density 
phase diagram of the hard-core bosonic model for $t^\prime/t=1$ and 
$t^\prime/t=2$. Calculations have been done on a $L=2 \times 50$ system.}
\end{figure}

\begin{figure}
\includegraphics[width=\columnwidth]{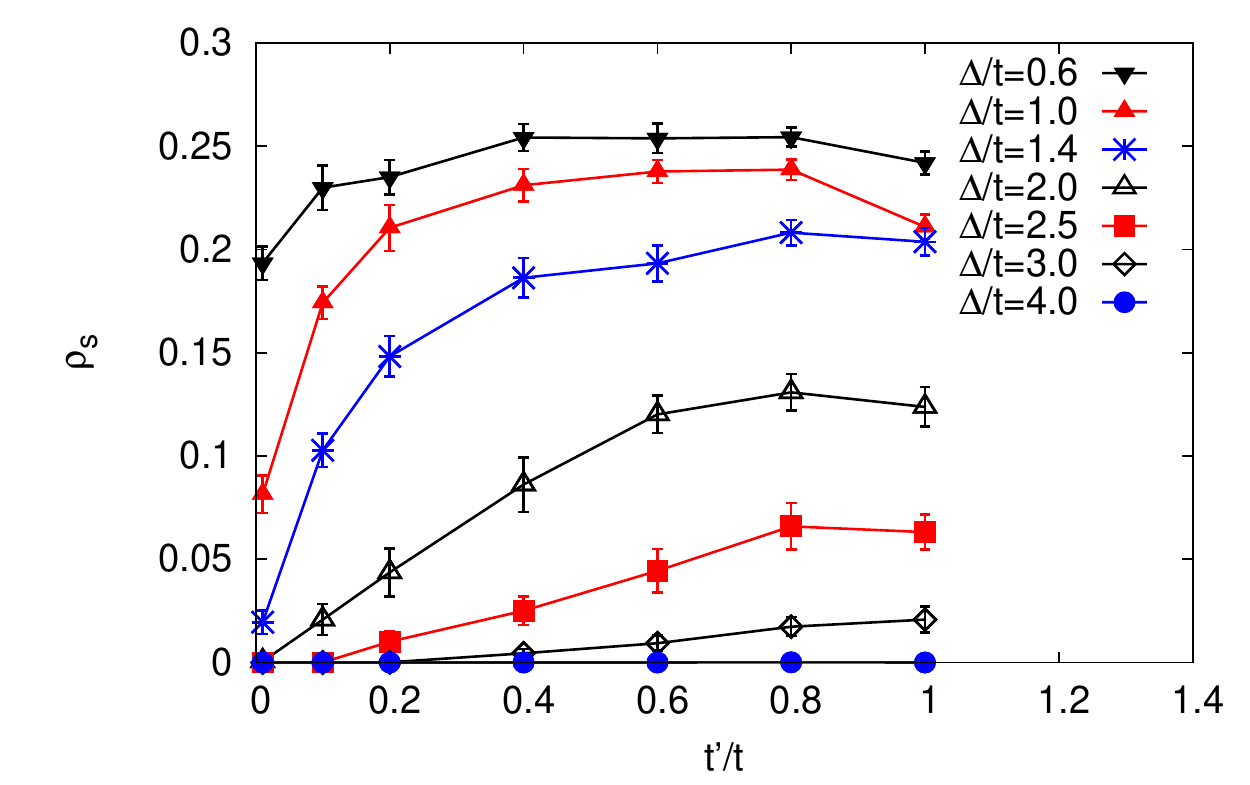}
\caption{\label{fig:smalltp}
(color on-line) Superfluid stiffness $\rho_s$ as a function of  $t^\prime/t$ 
for different $\Delta/t$ on a two-leg ladder with $L=2 \times 50$ sites. 
The density has been fixed to $n=0.4$.}
\end{figure}

\section{The disordered system}\label{sec:disordered}

\subsection{Low-density phase diagram}

Now we turn to the disordered case. Let us start by the low-density regime of 
the phase diagram at fixed inter-chain hopping as function of the density and 
disorder. In Fig.~\ref{fig:hcb}, we report our results of the superfluid 
stiffness as a function of disorder strength (for $t^\prime/t=1$) and density 
(for $t^\prime/t=2$); the low-density phase diagram is reported as well. 
We find that, for any finite disorder $\Delta/t$, the low-density phase is a 
Bose glass that turns superfluid above a critical density. The trivial Mott 
insulator with zero (or one) bosons per site is therefore always separated 
from the superfluid by the Bose-glass phase. This is a remarkable result since,
in a single chain with nearest-neighbor hopping only, hard-core bosons are 
equivalent to spin-less fermions, which Anderson localize for any density. 
Hence, in a two-leg ladder, hard-core bosons behave differently from spin-less 
fermions; while the latter ones remain always localized, the former ones show 
a superfluid phase stabilized by the inter-chain hopping, as it was predicted 
using bosonization and renormalization-group techniques.~\cite{orignac2} 
The idea is that, in a strictly 1D geometry with only nearest-neighbor hopping,
the statistics of the particles does not matter. However, whenever particles
may be interchanged (by non-strictly 1D paths) bosons can form a superfluid,
even in presence of disorder. We just mention that the same behavior holds 
also on a single chain with longer-range hopping. This scenario can be 
understood in very simple terms as follows. At very low fillings, the 
statistics of the particles does not matter so much and hard-core bosons, 
as free fermions, localize due to disorder, giving rise to the Bose glass. 
This is because the length over which the single-particle wave function extends
is short enough that the wave functions of two particles never overlap. 
As soon as the filling is increased, the particles get closer to each other 
and the single-particle wave functions begin to overlap. At this point the 
statistics of the particles starts to play a role. If the particles are 
fermions they will still be localized (in $D \le 2$), whereas bosons may 
stabilize a superfluid, as confirmed by our numerical simulations. 

In the range of values of inter-chain hopping that we have studied, the effect 
of a larger $t^\prime/t$ in the low-density phase diagram is to slightly 
reduce the superfluid response of the system, as can be also seen in 
Fig.~\ref{fig:hcb}. Although the actual thermodynamic value of the transition 
between the Bose glass and the superfluid may be rather different from the one
obtained by our calculations, because of strong size effects, the present 
results give a qualitative correct insight into the phase diagram. Finally, 
we would like to mention that the exact behavior of the transition line between
Bose glass and superfluid phases $\Delta_c(n) \propto n^\alpha$ is hard to be 
found by numerical calculations. Although an almost linear fit is found, 
i.e., $\alpha=1$, a different power-law cannot be excluded, as implied by
the arguments of Ref.~\onlinecite{falco}.

\subsection{The effect of the inter-chain hopping}

Here, we want to investigate the effect of the coupling $t^\prime$ on the 
otherwise insulating (Anderson localized) decoupled chains. Generally speaking,
for any value of disorder, a certain finite ratio $t^\prime/t$ is necessary to
drive the system into a superfluid phase, hence the system remains in the 
Bose-glass phase for small $t^\prime/t$. However, for small disorder, the 
localization length of the Bose glass is expected to be very large. This means
that on clusters that are accessible to numerical simulations it may be very 
hard to see the Bose glass region. This fact is indeed confirmed by our results
on the superfluid stiffness as a function of the inter-chain hopping, see 
Fig.~\ref{fig:smalltp}. Rapidly, as a small $t^\prime/t$ is introduced, a large
superfluid response is found for small disorder (e.g., $\Delta/t=0.6$ and $1.0$
in the figure). It is also observed that, by a further increase of the 
inter-chain hopping the superfluid stiffness reaches a maximum and then 
eventually decays, since for $t/t^\prime \to 0$ the system decouples in a 
collection of decoupled rungs (which are obviously not superfluid).
Moreover, as the disorder is increased, the superfluid response is suppressed,
until the system cannot attain superfluidity any longer and remains localized 
for any value of $t^\prime$.

\begin{figure}
\includegraphics[width=\columnwidth]{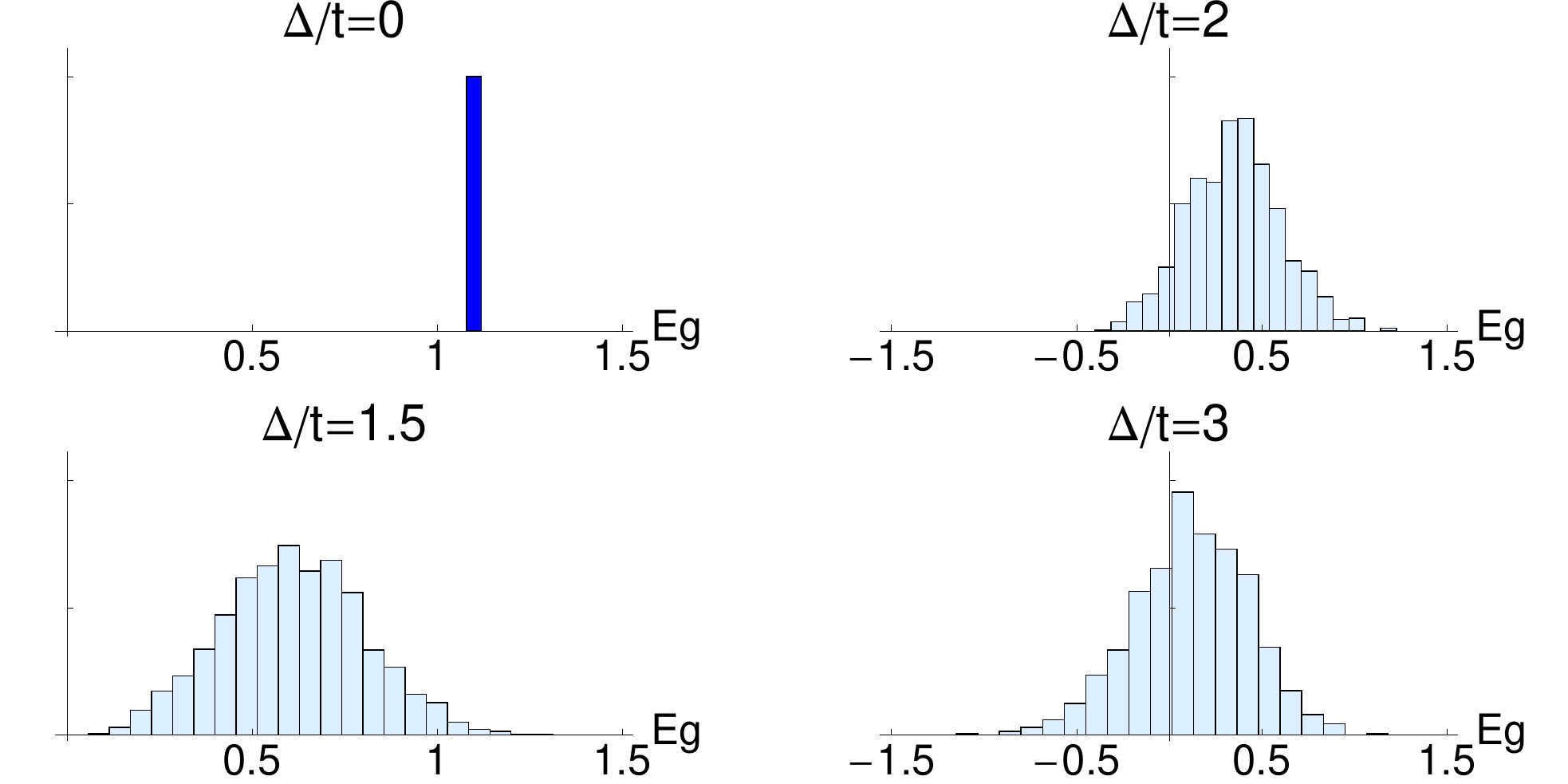}
\caption{\label{fig:distribs}
(Color on-line) Distribution $P(E_g)$ of the gap as function of disorder 
strength and fixed density $n=0.5$ and $t^\prime/t=2$ on a two-leg ladder 
with $L=2 \times 50$ sites. The clean gap is shown for comparison in the 
upper-left box as a blue bar.}
\end{figure}

\subsection{The rung-Mott phase in presence of disorder}

The effect of disorder on the rung-Mott phase at density $n=0.5$ system is 
now discussed. From general grounds, it is expected that the presence of 
disorder will fill the gap with localized states, so to induce a transition
to a gapless Bose-glass phase. In practice, given a ratio $t^\prime/t$, the 
rung-Mott insulator will survive up to a certain critical value of $\Delta/t$,
where the gap will be completely filled and the system will become 
compressible. This situation is similar to the one of the Bose-Hubbard 
model at integer fillings~\cite{fisher}, where a direct transition between
the Mott insulator and the Bose-glass phase is expected by decreasing the
ratio between the on-site repulsion $U$ and the disorder strength 
$\Delta$.~\cite{fisher,freericks,weichman,pollet} Also in our case of hard-core
bosons, we can make use of the argument based on the fact that, if $\Delta$ 
is larger than half of the energy gap of the clean insulator 
$E_{g}^{\rm clean}$, then the ground state must be compressible; otherwise 
the system is incompressible with a reduced gap given by 
$E_{g}=E_{g}^{\rm clean}-2\Delta$. In particular, for $t^\prime/t \gg 1$ we
have that $E_g^{\rm clean} \sim 2t^\prime$. Therefore, the gap will vanish 
around a critical value of disorder $\Delta_c \sim t^\prime$. 
These arguments should hold exactly only in the infinite system and large size
effects are expected because this transition is of the Griffiths type, i.e., 
driven by exponentially rare regions which are locally 
ordered.~\cite{griffiths} On the other hand, on any finite system the 
transition from the gapped to the compressible phase will appear at a larger 
$\Delta_c$, since these exponentially rare configurations will be hardly
sampled on finite clusters. 

We recently proposed~\cite{carrasqu} a method to alleviate the strong size 
effects that consists of computing directly the distribution probability of 
the gap
\begin{equation}
P(E_g) = \sum_{\alpha\beta}\,\delta \left (
E_g-\mu^{+}_\alpha + \mu^{-}_\beta \right ), 
\end{equation}
where $\mu^{\pm}_\alpha= \pm \left(E_{M\pm1}^\alpha-E_{M}^\alpha\right)$ 
($E_{M}^\alpha$ being the ground-state energy with $M$ particles on the  
realization $\alpha$ of disorder). This definition of the gap distribution 
is introduced because in disordered systems the gap can be overcome by 
transferring particles between two rare regions with almost flat disorder 
shifting the local chemical potential upward and downward. These exponentially
rare regions may be far apart in space and represent rare fluctuations 
(Lifshitz's tail regions), thus it is useful to imagine that a large system 
is made by several subsystems, each represented by a different disorder 
realization of our $L$-site cluster, and construct the gap by using the process
of taking one particle from region $\alpha$ to region $\beta$.~\cite{carrasqu}
If such processes are allowed at no energy cost, i.e., $P(0) \neq 0$, the 
corresponding system will be gapless. One could define an alternative estimate
of the gap as
\begin{equation}\label{mingap}
E_g^{{\rm min}}={\rm min}_{\alpha,\beta} 
\mid \mu^{+}_\alpha - \mu^{-}_\beta \mid,
\end{equation}
with all the disorder realizations $\alpha$ and $\beta$.

In Fig.~\ref{fig:distribs}, we show the distribution probability at half 
filling as function of disorder, $t^\prime/t=2$ and $L=2 \times 50$. For small 
values of disorder the gap survives, while for $\Delta/t=2$ the probability to 
find zero gap is finite, which we interpret as signalling zero gap in the 
infinite system and a Bose glass phase. We note that, when considering the case
of the rung-Mott insulator, this method performs a bit worse than in the case
of the Bose-Hubbard model (at integer fillings). Indeed, for this value of
the hopping parameters, we have that $E_g^{{\rm clean}}/t \simeq 1.11$, giving
rise to $\Delta_c/t \simeq 0.55$, which is much smaller than the value 
obtained by numerical simulations. However, we would like to mention that,
even though the finite-size analysis of $P(E_g)$ overestimates the actual
value of the transition, it gives a sizable improvement with respect to the
simple calculation of the average gap value in presence of disorder.
The precise determination of the critical point is well beyond any numerical 
calculations and, therefore, we take advantage of the fact that its
estimation can be done by using the criterion of 
Ref.~\onlinecite{fisher,freericks,weichman,pollet}, namely 
$\Delta_c=E_g^{{\rm clean}}/2$.

\begin{figure}
\includegraphics[width=\columnwidth]{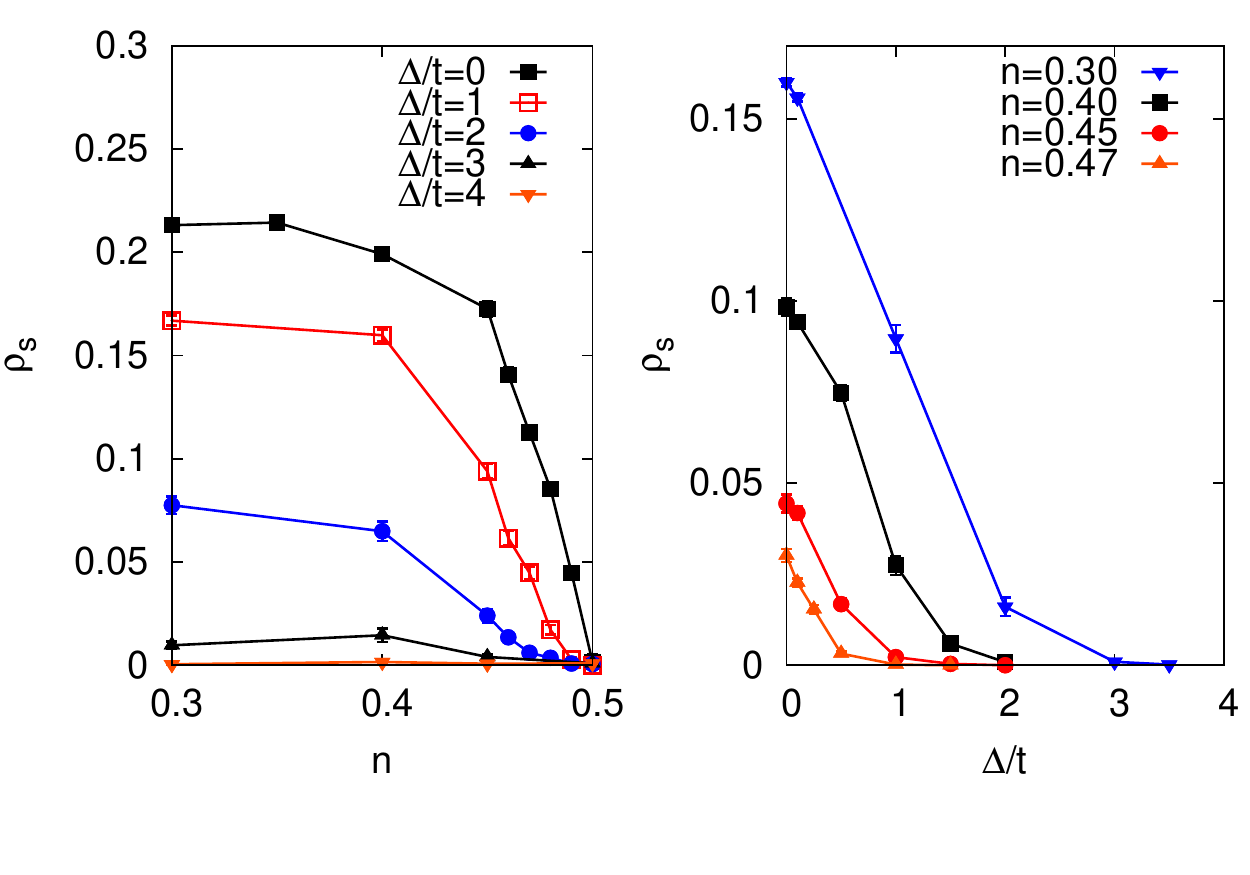}
\caption{\label{fig:largetp}
(Color on-line) Left panel: Superfluid stiffness $\rho_s$ as a function of 
the density and several values of $\Delta/t$ and fixed $t^\prime/t=2$. 
The stiffness of the clean system is also shown for comparison.
Right panel: Superfluid stiffness $\rho_s$ as function of the 
disorder bound for several densities close to half filling and fixed 
$L=2 \times 50$ and $t^\prime/t=10$.}
\end{figure}

\begin{figure}
\includegraphics[width=\columnwidth]{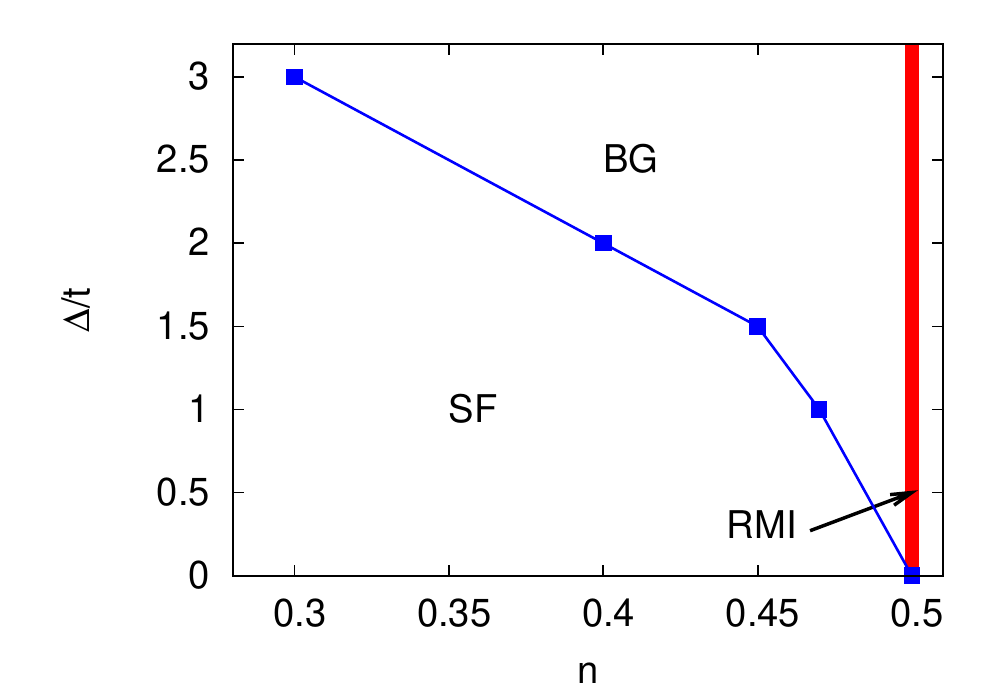}
\caption{\label{fig:tp10}
(Color on-line) Results of the $2 \times 50$ lattice for the phase diagram in 
the vicinity of the rung-Mott phase for $t^\prime/t=10$.}
\end{figure}

\subsection{Transition from the superfluid to the gapped phase}

In connection with the rung-Mott phase at half filling, we investigate the 
phase diagram in the vicinity of such a phase and address the question of 
whether it is possible to have a direct transition from the superfluid phase 
to the rung-Mott insulator as the density $n \to 0.5$, or there is always an 
intruding Bose glass phase. In this regard, whenever the gapped state is doped
with a few particles or holes such that their typical spacing will be large, 
those few carriers on top of the rung-Mott phase will effectively see a 
disordered background. Therefore, standard single-particle Anderson 
localization arguments apply and the system remains insulating by localizing 
those few carriers on the Lifshitz's tails that are in the Mott gap. 
By further increasing the density of particles (or holes), a superfluid is 
eventually formed. This simple single-particle argument implies the presence 
of an intervening Bose glass between the rung-Mott phase and the superfluid. 
We proceed to test this argument quantitatively. In Fig.~\ref{fig:largetp}, 
we report our numerical results for the superfluid stiffness at densities 
close to $n=0.5$ and $t^\prime/t=2$. Our data is consistent with a transition 
driven by density from the superfluid phase through the Bose glass to finally 
end up with the rung-Mott insulator. For example, for $\Delta/t=1$, the 
superfluid stiffness appears to vanish just before $n=0.5$, however the region
in which the Bose glass takes place is very small. For $\Delta/t=2$ the 
rung-Mott insulator has already been wiped out by the effect of disorder, 
as observed in Fig.~\ref{fig:distribs}, and, therefore, this issue cannot be 
addressed.

By considering a larger value of $t^\prime/t$ we have two advantages: first,
the Mott gap is larger, such that the gapped phase is more robust, second, the 
localization due to disorder is expected to be enhanced, thus, enlarging the 
Bose glass region at $n<0.5$ (or $n>0.5$). These facts enable us to provide 
further evidence in favor of an intervening Bose glass in between the 
rung-Mott and the superfluid as follows. We have performed simulations with 
a rather large $t^\prime/t=10$. For such a value of the hoppings, we have 
that $E_{g}^{\rm clean} \simeq 17.31$, such that the transition from 
the gapped to the compressible (Bose-glass) phase is argued to occur at a 
$\Delta_c \simeq 8.65$. Therefore, the system is expected to be gapped for 
$\Delta \lesssim 8.65$. In Fig.~\ref{fig:largetp}, we present our results for
the superfluid stiffness as function of the disorder strength for several 
densities close to half filling. From these calculations we can easily see 
that, as the density approaches $n=0.5$, the critical point where the 
stiffness vanishes gets smaller, leaving room for a large Bose-glass phase in 
between the superfluid and the rung-Mott insulator. Given our results, we can 
draw the phase diagram for densities close to the rung-Mott phase, see
Fig.~\ref{fig:tp10}. Notice that the large value of $t^\prime/t$ (and therefore
the clean Mott gap) ensures the existence of a truly gapped state at half
filling. All together, we can make the safe statement that the transition
between the superfluid and the rung-Mott phases is not direct, but through
an intervening Bose-glass state.

\begin{figure}
\includegraphics[width=\columnwidth]{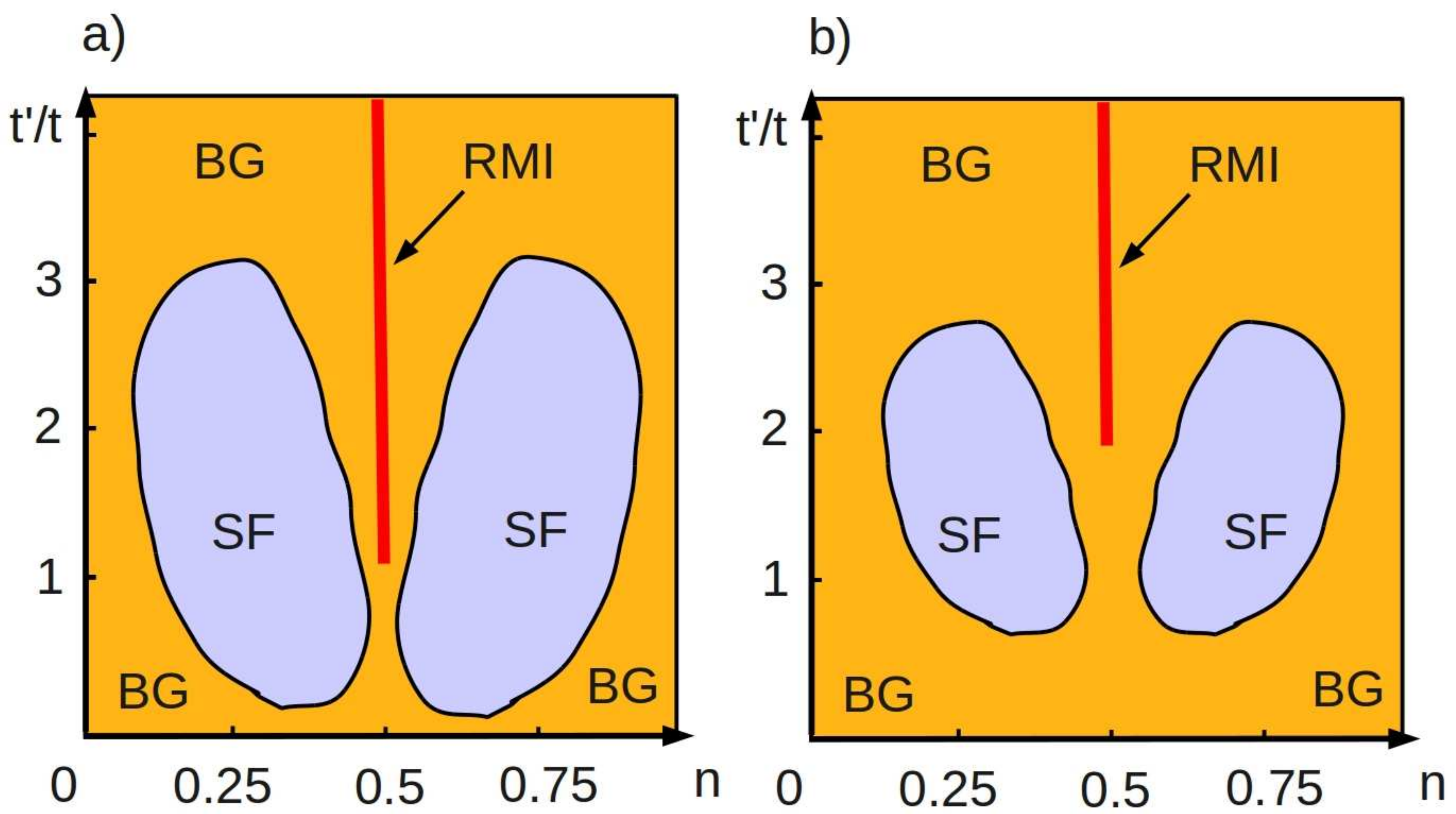}
\caption{\label{fig:phasediag}
(Color on-line) a) Tentative zero-temperature phase diagram of hard-core bosons
on the two-leg ladder system. b) Same phase diagram in a) but for a larger 
value of disorder $\Delta/t$.}
\end{figure}

\section{Conclusions}\label{sec:concl}

We have studied hard-core bosons on disordered two-leg ladders by using both
numerical techniques and analytical arguments borrowed from similar
problems in Bose-Hubbard models. We have shown that the zero-temperature
phase diagram is rather rich and contains different phases; apart from the
trivial Mott insulators at $n=0$ and $1$, that are totally frozen due to
the hard-core constraint, we found superfluid, Bose-glass, and rung-Mott 
phases. This contrasts the case of spin-less fermions, where no metallic phases
are possible and Anderson localization takes place for any density $n \ne 0.5$
at finite $\Delta$. A final sketched phase diagram, based upon our results,
is reported in Fig.~\ref{fig:phasediag}. In the case of no disorder, i.e.,
$\Delta=0$, the superfluid phase pervades the phase diagram for all densities
$0<n<0.5$ and $0.5<n<1$ and all $t^\prime/t \ne \infty$. When considering a
finite disorder strength, the superfluid shrinks and a Bose-glass phase 
appears. Most importantly, the transition between the Mott and the rung-Mott 
phases and the superfluid ones is never direct, like in the Bose-Hubbard 
model.~\cite{fisher,freericks,weichman,pollet} 

\acknowledgments

We thank N. Laflorencie for useful discussion.
F.B. thanks T. Giamarchi for interesting discussions in Santa Barbara, during 
the program ``Disentangling Quantum Many-body Systems: Computational and 
Conceptual Approaches''. F.B. wants to acknowledge the fact that this research
was supported in part by the National Science Foundation under the Grant 
No. NSF PHY05-51164.

\appendix
\section{Basic bosonization formulas}

There are several papers that discuss at length the harmonic-fluid 
representation of bosonic lattice Hamiltonians following the seminal work by 
Haldane.~\cite{haldane} Nevertheless, we believe that it is worth listing some
useful formulas, referring the interested readers to existing literature for 
further details.~\cite{haldane,giamarchi,cazalilla}  

In the long-wavelength limit, the boson density and creation operator on  
each chain $\alpha=1,2$ can be written as
\begin{eqnarray*}
\rho_\alpha(x) &=& \bigg(\rho_0 +\frac{1}{\pi}\nabla\phi_\alpha(x)\bigg) 
\sum_{m=-\infty}^\infty \mathrm{e}^{i2m\Big(\phi_\alpha(x)+\pi\rho_0 x\Big)}, \\
\psi^\dagger_\alpha(x) &=& \sqrt{\rho_\alpha(x)}\;\mathrm{e}^{i\theta_\alpha(x)} \\
&\sim& \mathrm{e}^{i\theta_\alpha(x)} 
\sum_{m=-\infty}^\infty \mathrm{e}^{i2m\Big(\phi_\alpha(x)+\pi\rho_0 x\Big)},
\end{eqnarray*} 
where $\rho_0$ is the average density and the two fields $\phi(x)$ and 
$\theta(x)$ satisfy 
\begin{equation}
\Big[\phi_\alpha(x),\nabla\theta_\beta(y)\Big] = i\pi \delta_{\alpha\beta}\,\delta(x-y).
\end{equation} 
In the case of hard-core bosons, it could be useful to define $\phi(x)$ and 
$\theta(x)$ in terms of right (R) and left (L) chiral fields:~\cite{GNT}
\begin{eqnarray}
\phi_\alpha(x) &=& \frac{1}{2}\Big(\phi_{\alpha R}(x) + \phi_{\alpha L}(x)\Big),\\
\theta_\alpha(x) &=& \frac{1}{2}\Big(\phi_{\alpha L}(x) - \phi_{\alpha R}(x)\Big).
\end{eqnarray}
The dimension $d$ of each operator $\Delta(x)$, defined through 
$\langle \Delta(x)\,\Delta(0)\rangle \sim x^{-2d}$, can be easily evaluated 
by recalling that 
\[
\langle \mathrm{e}^{i\gamma\phi_R(x)}\,\mathrm{e}^{-i\gamma\phi_R(0)}\rangle 
\sim 
\langle \mathrm{e}^{i\gamma\phi_L(x)}\,\mathrm{e}^{-i\gamma\phi_L(0)}\rangle 
\sim \left( \frac{1}{x}\right)^{\gamma^2}.
\]
In the two-leg ladder it is convenient to introduce the symmetric (s) and 
anti-symmetric (a) combinations $\phi_s = \Big(\phi_1+\phi_2\Big)/\sqrt{2}$ and 
$\phi_a = \Big(\phi_1-\phi_2\Big)/\sqrt{2}$, respectively (and seemingly for 
$\theta_s$ and $\theta_a$). It follows that the inter-chain hopping becomes
\begin{eqnarray*}
\psi^\dagger_1(x)\,\psi^\dagga_2(x) &\sim& 
\mathrm{e}^{-i\sqrt{2}\,\theta_a(x)}\;
\sum_{m,n}\, \mathrm{e}^{i2\pi\rho_0 (m+n) x} \times \\
&\times& \mathrm{e}^{i\sqrt{2}\,(m+n)\,\phi_s(x)}\;\mathrm{e}^{i\sqrt{2}\,(m-n)\,\phi_a(x)}.
\end{eqnarray*}  
The first term in the right hand side is the most relevant one and opens a gap
in the anti-symmetric sector such that $\theta_a$ acquires a finite average 
value while $\phi_a$ has exponentially decaying correlations. It follows that 
the leading operator generated by the inter-chain hopping is 
\begin{equation}\label{app-hopping}
\psi^\dagger_1(x)\,\psi^\dagga_2(x) \sim 
\mathrm{e}^{-i\sqrt{2}\,\theta_a(x)}\;\bigg[ 
1 + 2\cos\Big(\sqrt{8}\,\phi_s(x) +4\pi\rho_0 x\Big)\bigg].
\end{equation}
In other words, $t^\prime$ not only gaps the anti-symmetric sector but also 
generates an umklapp scattering in the symmetric channel that becomes 
marginally relevant at half filling, where $4\pi\rho_0=2\pi$. 
It is just this umklapp that is responsible for the appearance of the rung-Mott
insulator at any finite $t^\prime \ll t$. In addition, being marginally 
relevant, it opens a gap in a Berezinskii-Kosterlitz-Thouless fashion, which 
is hard to detect numerically. Following Ref.~\onlinecite{giamarchi}, one 
finds that disorder gives rise to a fluctuating umklapp
\begin{equation}\label{app-disorder}
{\cal H}_{\rm umklapp} = 
\int dx\, \xi(x)\,\cos\Big(\sqrt{8}\,\phi_s(x)\Big),
\end{equation}
with $\overline{\xi(x)\xi(y)} = u^2\,\delta(x-y)$ in case of a Gaussian noise.
Therefore, at half filling the coupling constant of the umklapp has a finite 
value plus a fluctuating one $\xi(x)$; therefore, if the latter one is small,
the gap is on average finite (the Mott phase), while, for $u$ above a certain 
threshold, the gap is washed out by disorder, leading to the Bose glass. 

Away from half filling, we should keep into account a renormalization of the 
symmetric sector that can be parameterized by a Luttinger liquid parameter 
$K_s$ through~\cite{haldane,GNT,giamarchi}
\[
\phi_s \rightarrow \sqrt{K_s}\;\phi_s,\qquad 
\theta_s \rightarrow \sqrt{\frac{1}{K_s}}\;\theta_s. 
\]
Since the full density 
\[
\rho(x)=\rho_1(x)+\rho_2(x) = \frac{\sqrt{2}}{\pi}\,\nabla \phi_s(x) 
\rightarrow \frac{\sqrt{2K_s}}{\pi}\,\nabla \phi_s(x),
\]
$K_s$ can be easily extracted by the static structure factor in momentum space 
\[
\langle \rho(q)\,\rho(-q)\rangle = 2K_s\,\frac{q L_x}{2\pi},
\]
where $L_x$ is the number of sites per chain.
When $\rho_0 \ne 1/2$, the umklapp scattering in Eq.~(\ref{app-hopping}) 
ceases to play a role and what survives is just the disorder-generated 
term~(\ref{app-disorder}). Conventional scaling arguments predict that such a 
term is relevant when $K_s<3/4$.~\cite{giamarchi} This would suggest that for 
$K_s>3/4$ a superfluid phase is stable, otherwise disorder is relevant and the
Bose glass occurs. According to the theory of the commensurate-incommensurate 
transition in 1D, we expect in the clean case that, as the density $\rho_0$ 
approaches half filling $K_s \to 1/2$, which would imply that disorder becomes
relevant already before the Mott transition is approached in density. However,
even if the density $\rho_0 \to 0$ we should expect $K_s\to 1/2$. Therefore, 
whatever is the behavior close to half filling, it must be qualitatively the 
same also close to zero filling. We know that, at very low density, bosons 
localize in the Lifshitz's tails and superfluidity can arise only when the 
localization length becomes of the order of the interparticle distance. This 
also suggests that a harmonic-fluid representation is likely inadequate at 
low density, hence that the scaling criterium $K_s>3/4$ for the appearance of 
superfluidity may not work. Seemingly, the same argument must apply close to 
half filling, so that it must not be surprising that the phase boundary between 
the superfluid and the Bose glass goes smoothly and almost linearly to zero, 
see Fig.~\ref{fig:largetp}.

\end{document}